\documentclass[letterpaper]{article}
\usepackage{aaai18}
\usepackage{amsmath}

\usepackage{times}
\usepackage{helvet}
\usepackage{courier}
\usepackage{graphicx}
\usepackage{tabularx,ragged2e}

\begin{document}
%
\title{Life in the {\it Matrix}\thanks{A metaphor taken from the hit movie Matrix. The work is to appear at ICWSM 2018.}: Human Mobility Patterns in the Cyber Space}

\author{Tianran Hu, Jiebo Luo \\
Department of Computer Science \\
University of Rochester \\
\{thu, jluo\}@cs.rochester.edu \\
\And Wei Liu \\
AI Lab\\
Tencent Inc.\\
wliu.cu@gmail.com\\
}
\maketitle

\begin{abstract}
With the wide adoption of the multi-community setting in many popular social media platforms, the increasing user engagements across multiple online communities warrant research attention. In this paper, we introduce a novel analogy between the movements in the cyber space and the physical space. This analogy implies a new way of studying human online activities by modelling the activities across online communities in a similar fashion as the movements among locations. First, we quantitatively validate the analogy by comparing several important properties of human online activities and physical movements. Our experiments reveal striking similarities between the cyber space and the physical space. Next, inspired by the established methodology on human mobility in the physical space, we propose a framework to study human ``mobility'' across online platforms. We discover three interesting patterns of user engagements in online communities. Furthermore, our experiments indicate that people with different mobility patterns also exhibit divergent preferences to online communities. This work not only attempts to achieve a better understanding of human online activities, but also intends to open a promising research direction with rich implications and applications. 
\end{abstract}

\section{Introduction}

Understanding human activities in online communities not only is the key to computational sociology research~\cite{ren2007applying,zhu2014impact}, but also can offer valuable guidance to the design of online systems~\cite{kraut2012building}. Many popular platforms, such as Reddit, 4chan, and StackExchange, adopt the setting of multiple communities for user engagement. Much work has been done on human activities across communities, including user exploration and participation patterns in more than one community~\cite{tan2015all,zhang2017community}, and user loyalty under the multi-community setting~\cite{hamilton2017loyalty}. However, these studies usually focus on specific aspects of online activities, while a comprehensive and general-purpose framework for studying multi-community activities is still lacking. To overcome the problem, we introduce a novel analogy between human online activities and offline physical movements in this paper. Given the rich body of research on human mobility patterns~\cite{barbosa2017human}, such an analogy allows us to borrow the existing approaches and frameworks for analyzing movements in the physical space and apply them to online scenarios with necessary adaptation. In this work, we first validate this analogy, and design experiments to reveal striking quantitative similarities between human online activities and offline movements. We then propose a frameworks for studying users of online platforms, and uncover interesting activity patterns across communities.

We draw a strong analogy between the cyber space (i.e. online multi-community platforms such as Reddit)\footnote{We use ``cyber space'' and ``online multi-community platforms'' interchangeably in this paper for the simplicity of narration. Please note that ``cyber space'', as sometimes used as a metaphor to the whole Internet, is a wider concept than online platforms.} and physical space. The communities on the online platforms is then treated as the \textit{locations} in the cyber space. The intuition of the analogy arises from two aspects: 
\begin{itemize}
\item The activities in the cyber space resemble the movements in the physical space -- people move from one community to another on the platforms, explore new communities, and regularly visit communities with which they are familiar, as do they with physical locations. 
\item The locations in the two spaces (i.e. communities and places) share important similarities -- some locations are popular and thus gain large amounts of visitors; some are niche and attract specific groups of visitors; some are private where only authorized visitors have access (e.g. home and private subreddit). 
\end{itemize}
Without causing ambiguity, in this paper we may use the \textit{movements} across locations in the cyber space to refer to the activities across online communities.

We quantitatively validate our analogy by comparing the properties of movements in the cyber and physical spaces from three representative and progressive aspects:  1) at a coarse granularity, we first compare the overall visit distributions in the two spaces; 2) at a fine granularity, we then study the visit behavior of individuals; and 3) considering time factors, we further investigate the temporal properties of both cyber and physical movements. The results reveal striking similarities between the movements in the two spaces. For example, the number of visits to an online community is found to fall into the same distribution of the number visits to a physical location. Moreover, it is shown that individuals visit locations in both spaces following random-walk behavior. More strikingly, the Zipf's law of the movements in the physical space also applies to the movements in the cyber space. In terms of temporal property, we observe an almost identical returning pattern in both spaces -- people tend to return to specific communities/locations on a daily basis. Also, a temporally complementary relation between the two spaces is uncovered -- human mobility level is high in the cyber space when the level is low in the physical space, and vice versa.

The analogy implies a possibility of modelling the movements in the cyber space using the existing approaches for studying physical movements. However, one of the differences between the two spaces is that distance is not universally defined in the cyber space~\cite{hessel2016science,pavalanathan2017multidimensional}. Instead of arbitrarily defining the distance concept, we focus on entropy-based approaches that do not require an explicit measurement of distance. Entropy and its variants have been widely used for analyzing the randomness and predictability of individual mobility~\cite{song2010limits}, measuring diversity of visitors to a location~\cite{cranshaw2010bridging}, and so on. Following this line of research, we study human mobility in the cyber space. 

We first study the randomness of human mobility in the cyber space. We discover that mobility randomness varies significantly across people. Some people evenly distribute their visits to many communities, and therefore exhibit high mobility randomness. In contrast, some people only visit a very limited number of communities, and exhibit very low mobility randomness. We then investigate if human mobility in the cyber space evolves over time. By comparing different stages of a user's online lifespan, our experiments reveal three interesting mobility patterns in the cyber space: 1) concentrating on a limited number of communities throughout the whole lifespan; 2) exploring many communities in the early stage, but stopping exploring and concentrating on only a few later on; and 3) focusing on a few communities in the early stage, and starting to explore more and more communities later on. Furthermore, we observe that people of the concentrated pattern are more likely to visit communities of specific topics and smaller sizes, while people of both two exploratory patterns prefer communities of general topics and larger sizes.

The main contributions of this paper are threefold:

\begin{itemize}
    \item We introduce a novel analogy between the cyber space and the physical space. This work is intended to first achieve a better understanding of human online activities, and more importantly, initialize a promising new research direction.
    
    \item We quantitatively examine the validity of the analogy between the cyber space and the physical space. Our experiments reveal striking similarities between the two spaces.
    
    \item We investigate the individual mobility patterns in the cyber space. Our results reveal three major patterns of online activities, as well as the different preferences to online communities by the people of the three patterns.
    
\end{itemize}

\section{Related Work}

\subsection{Online Communities}

Rotman et al. suggest that users interact and share purpose in online groups, and therefore, form sociological communities~\cite{rotman2009community}. Since then there has been a rich literature studying online communities from various aspects, such as the styles and evolution of communities~\cite{tran2016characterizing,lin2017better}, differences between communities~\cite{hessel2016science}, and so on. The interaction between users and communities has also been investigated. For example, \cite{danescu2013no} studies the changes of linguistic style in online communities, and uncovers the interesting evolution of the interaction patterns between users and communities.

Given the wide adoption of the multi-community setting on online platforms, more and more attention has been paid to human activities across online communities~\cite{chan2010decomposing,pavlick2016empirical}. Zhu et al. report that user participation in multiple communities benefits the survival of communities~\cite{zhu2014impact}. Ren et al. study the common identities within different communities, and suggest the ``bonds'' between online communities~\cite{ren2007applying}. Furthermore, it is reported that community characteristics can affect the user engagement pattern in a community.~\cite{hamilton2017loyalty}. For example, a community with a distinctive and dynamic identity is not only more likely to retain users, but also creates a larger ``cultural'' gap between senior members and newcomers~\cite{zhang2017community}. Tan et al. study the involvements of users in more than one communities~\cite{tan2015all}. This work suggests that instead of gradually settling down in previously visited communities, online users keep exploring new but less popular communities.


\subsection{Human Physical Mobility Patterns}

Much research work has been devoted to human mobility in the physical space on various topics such as individual and group level mobility patterns~\cite{de2013unique,simini2012universal}, the temporal-spatial properties of human mobility~\cite{hu2017tales}, and the relation between individual mobility and social connections~\cite{cho2011friendship}. Many studies in modelling human mobility suggest the relation between random-walk models and the movements in the physical space~\cite{gonzalez2008understanding}. For example, Song et al. model people in the physical space as randomly moving objects, and propose methods to predict human mobility~\cite{song2010modelling}. Many interesting temporal patterns of human mobility have also been discovered in previous work. For example, \cite{cheng2011exploring} indicates that people return to specific locations on a daily basis, and \cite{noulas2011empirical} reports the different mobility levels of people on weekdays and weekends. Without modelling the distance in the space, a number of entropy-based approaches are proposed to investigate the randomness of human mobility~\cite{cranshaw2010bridging,smith2014refined}. For example, Song et al. study human moving trajectories using entropy, and discover a high predictability in human mobility~\cite{song2010limits}.

\section{Data Collection and Preprocessing}

We collect our data from Reddit, which was launched in 2005 and is now one of the most visited websites in the world\footnote{www.en.wikipedia.org/wiki/Reddit}. Due to its high popularity, long time span, and almost complete data availability\footnote{Reddit data is made publicly available, and free for download at www.reddit.com/3bxlg7}, Reddit has been used as the data source in many previous studies~\cite{hamilton2017loyalty,zhang2017characterizing}. Reddit is organized into thousands topic-based communities (subreddits), and users are allowed to join any communities at will (except for private subreddits). Such a multi-community setting makes Reddit ideal for our study. 

We download all the posts on Reddit from the website's inception on Dec 2005 to Dec 2016. The dataset contains 2.9 billion posts sent by 21 million users on 430 thousand subreddits. Besides text content, each post in the dataset is associated with many other types of information, such as user ID, community ID, time stamp, and so on. We first filter out 0.3 billion posts sent by deleted users (denoted by a user ID of ``[deleted]'' in the data). We also remove the posts by non-human accounts. To detect non-human accounts (bots), we first collect the users that post with an abnormally high frequency (50 thousand+ posts), and take them as possible examples of non-human accounts. We observe the user IDs of these accounts, and summarize a list of terms that frequently occur in these user IDs, such as ``-bot'', ``\_transcriber'', and ``Moderator''. We take the accounts that contain at least one of these terms in their IDs as non-human accounts, and remove all the posts sent by these accounts. In total, we filter out 35 million posts sent by 28 thousand user accounts of this kind. After the data cleaning, we further process the data to extract the visit history of each user. To be specific, the visit history of a user records all the communities the user has visited in the chronological order. In other words, visit histories are users' ``trajectories'' in the cyber space. Our work investigates the mobility patterns across communities by focusing on user visit histories. Different slices of user visit histories may be used for facilitating the different problems studied in this paper. We will specify the data used for each problem in the corresponding section.

\begin{figure*}
\centering
\includegraphics[width=2\columnwidth]{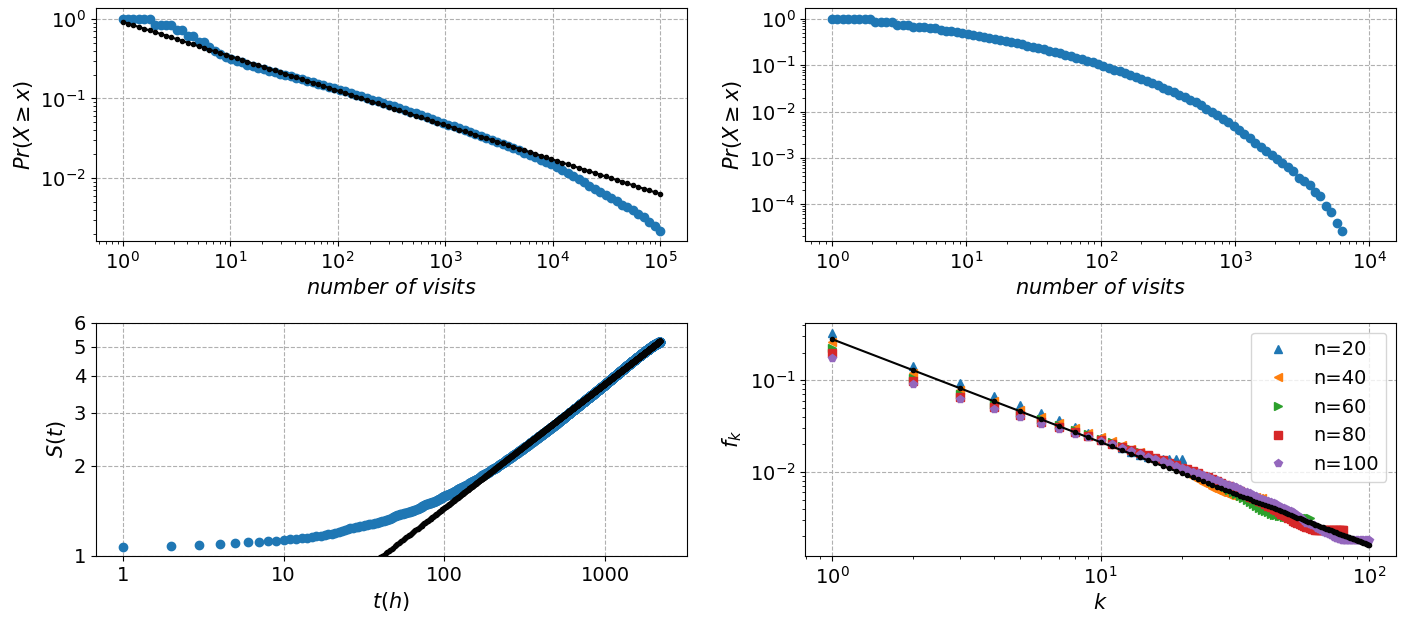}
\caption{(a) Complementary Cumulative Distribution Function (CCDF) of the number of visits to an online community. (b) CCDF of the number of visits per user. (c) Distribution of the number of visited distinct communities $S(t)$. (d) Distributions of the frequency of the $k$th most visited community $f_k$ for different $S$ values.}~\label{fig:CCDF}
\end{figure*}

\section{Analogy between the Two Spaces}

We first validate the analogy between the cyber space and the physical space. To be specific, we study if the properties of human mobility discovered in the physical space still hold true in the cyber space. On the data collected from Reddit, we conduct the experiments that are originally designed for studying human physical mobility, and compare our results with the conclusions reported in the previous work. Most work on physical mobility is based on data collected in time spans of several months~\cite{barbosa2017human}. To align with the previous work, we collect a data slice containing all the posts on Reddit from January to March 2016. This data slice contains 169 million posts sent by 4.6 million users on 161 thousand subreddits (after data cleaning). All the experiments in this section are conducted on this data slice.

\subsection{Distributions of Visits}

At a coarse granularity, we first compare the distributions of the number of visits in both spaces. \cite{noulas2011empirical} reports the complementary cumulative distribution function (CCDF) of the number of visits to a physical location, as well as the CCDF of the number of visits per user. Both distributions reportedly have heavy tails. Moreover, the trend of the number of visits to a physical location follows a power-law distribution (a straight line in log-log scale). We plot the CCDF of the number of visits to an online community, as well as the CCDF of the number of visits per user in Figure~\ref{fig:CCDF} (a) $\sim$ (b). The plot shows that both distributions of visits in the cyber space exhibit the same trends as in the physical space. To be specific, both distributions computed from online data also have heavy tails. Moreover, the trend of the number of visits to an online community follows a power-law distribution (Figure~\ref{fig:CCDF} (a)).

The same trends of the distributions in the two spaces imply the similarities between communities and physical locations, as well as the similarities between the activities across communities and the movements across locations. The heavy tailed distribution of the visit amount to an online community reveals that, similar to physical locations, only a few communities receive a large number of visits, and a higher number of communities have only few visits. Meanwhile, the same distributions of visits per user suggest that, in both cyber and physical spaces, a small number of people contribute a large amount of visits, while the number of visits of most people is low.

\subsection{Human Visit Behavior}
At a fine granularity, we compare human visit behavior in the two spaces. Much work on human mobility in the physical space suggests the relation between random-walk models and the physical movements of individuals~\cite{gonzalez2008understanding,castellano2009statistical}. Human physical mobility reportedly exhibits two important quantitative characteristics of random-walk behavior~\cite{song2010modelling}:

1) the number of distinct locations visited by a user, denoted by $S(t)$, follows 
\begin{align}
  S(t) \sim t^\mu  
\end{align}
where $t$ is the time the user spent in the space. 

2) the frequency $f_k$ of the $k$th most visited location of a user follows Zipf's law, and formally, 
\begin{align}
f_k \sim k^{-\zeta}    
\end{align}
The first characteristic describes the exploration of people in the physical space. The parameter $\mu$ is estimated to be smaller than 1 from physical mobility data, indicating a decreasing tendency of  the users to visit new locations through time. The second characteristic indicates that the visits of users are distributed very unevenly, with most visits paid to a few most visited locations. We follow the exact steps suggested in~\cite{song2010modelling} to investigate if these two characteristics still apply to the cyber space.

\subsubsection{Distribution of Visited Locations}
For the first characteristic, we first extract the visit histories of all the users from the three month data slice. For a user, we split the visit history into hours, and compute the number of distinct communities visited by the end of each hour, denoted by $s(t)$. Then $S(t)$ is empirically computed as the average of $s(t)$ of all the users. We plot the relation between $S(t)$ and $t$ in Figure~\ref{fig:CCDF} (c). The result reveals that in the cyber space the relation $S(t) \sim t^\mu$ also holds, suggesting a similar exploration pattern of people in the two spaces. Furthermore, the parameter $\mu$ is estimated to be 0.4 in our experiment, which is smaller than the estimation in the physical space ($0.6\pm0.02$). This indicates that the tendency of user visiting new communities in the cyber space also decreases over time. Moreover, the decreasing tendency is faster in the cyber space than in the physical space.

\subsubsection{Zipf's Law}
We then validate the Zipf's law in the cyber space. From the data slice, we select the users who visited $S$ unique communities. Different values of $S$ are experimented as suggested in~\cite{gonzalez2008understanding,song2010modelling}. We then sort the communities a user visited according to their visit frequencies. The frequency $f_k$ is computed as the average of the visit frequencies to the $k$th most visited communities of all the users. Please note that $k$ ranges from 1 (most visited) to $S$ (least visited). The relation between $f_k$ and $k$ is plotted in Figure~\ref{fig:CCDF} (d). The result shows that the Zipf's law $f_k \sim k^{-\zeta}$ also applies in the cyber space. More strikingly, the parameter $\zeta$ is estimated to be 1.12 from our data, which is very close to the results estimated on physical mobility data ($1.2\pm0.1$). This indicates that people distribute their visits unevenly in a very similar fashion in both the cyber space and physical space.

Our experiments reveal the two characteristics also apply to the movements in the cyber space. The results suggest that individuals' trajectories across online communities can also be described using random-walk models. This further implies the similarity between human mobility in the cyber space and the physical space.

\begin{figure}
\centering
\includegraphics[width=1\columnwidth]{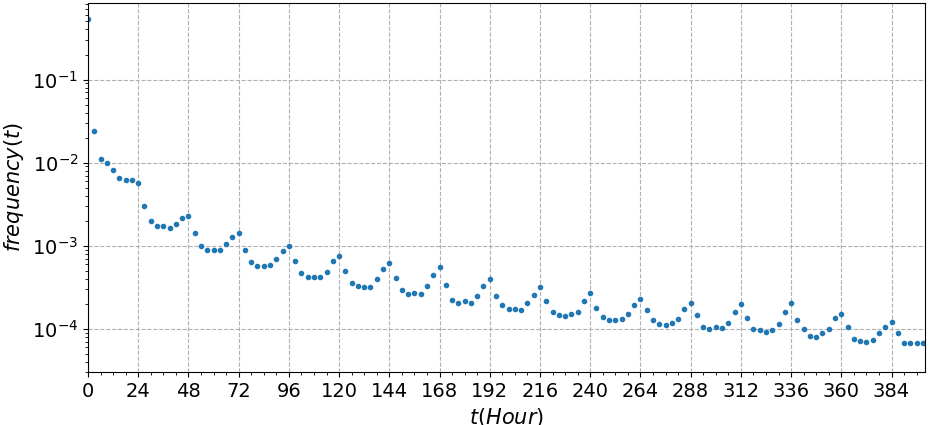}
\caption{Distribution of the returning probability to online communities.}~\label{fig:returning}
\vspace{-2em}
\end{figure}

\subsection{Temporal Properties}

Next, we validate our analogy for two well studied temporal properties of human physical mobility: distribution of returning probability~\cite{gonzalez2008understanding,cheng2011exploring} and hourly mobility levels~\cite{noulas2011empirical,hu2016home}. Returning probability measures the periodic patterns of human mobility. To be specific, the returning probability is the probability that a user returns to a location that the user visited $t$ hours before. In the physical space, the distribution of the returning probability, although having an overall decreasing tendency over time, increases sharply every 24 hours. This indicates a daily pattern of human mobility -- people return to specific locations on a daily basis. Zooming into the hourly level, previous work reports a high level of mobility during the daytime and a low level during the nighttime. Furthermore, researchers have observed two peaks in the mobility level around 9am and 7pm on weekdays, matching rush hours in the morning and happy hours in the evening, respectively. As for weekends, human mobility increases rapidly in the morning, and stays at a high level from 2pm to 9pm.

\subsubsection{Returning Probability}
We first compute the distribution of returning probability in the cyber space. For each community a user visited, we record the time gap between every two consecutive visits to the community (i.e. returning time). We then collect the returning time of all the users, and compute the probability of returning to a community in the $t$th hour. The distribution is plotted in Figure~\ref{fig:returning}. Quite interestingly, we observe an almost identical distribution of returning probability as in the physical space -- the probability has an overall decreasing tendency, but increases sharply every 24 hours. It indicates that people return to specific online communities also on a daily basis. 

\subsubsection{Hourly Mobility Level}
\begin{figure}
\centering
\includegraphics[width=1\columnwidth]{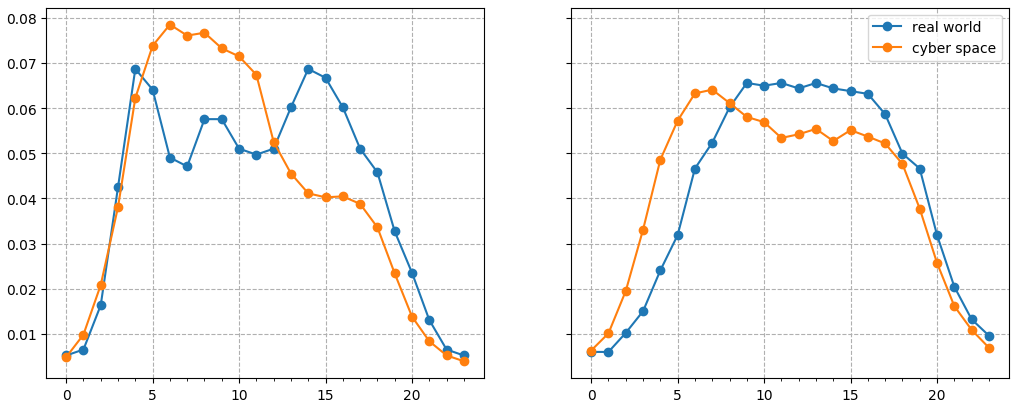}
\caption{Mobility levels in the cyber space over weekdays (left) and weekends (right). Please note that the two subplots share the same y-axis for a better comparison between weekdays and weekends. The values of frequency of visits in the physical space are learned from~\cite{noulas2011empirical}.}~\label{fig:weekdays}
\vspace{-1em}
\end{figure}

We then compute the mobility level in the cyber space. One of the requirements for computing the hourly mobility level is to know the clock time of each movement. However, time stamps on Reddit are recorded in UTC time while no time zone information is available. Fortunately, there are many subreddits on specific cities (e.g. /r/nyc\footnote{On Reddit, a subreddit is denoted as ``/r/'' + an unique name.}). Since the topics in a city-specific subreddit are mostly related to the life in the city, we can assume that most posts in such subreddits are sent by the city residents. Therefore, given the time zone of a city, we are able to compute the local time for each post in the city specific subreddit. Following this idea, we first collect the posts from several popular city-specific subreddits such as /r/nyc, /r/boston, /r/LosAngeles, and so on. We then convert the time stamp of each post to the local time of the corresponding city. The mobility level of one hour in the cyber space is computed as the percentage of posts in the hour. The mobility levels on both weekdays and weekends in the cyber space are shown in Figure~\ref{fig:weekdays}. We also plot the mobility levels in the physical world as reported in~\cite{noulas2011empirical} for a better comparison. 

Similar to the physical space, we observe that the mobility level in the cyber space is much higher during the daytime than during the nighttime. Also, on weekdays the mobility level in the cyber space varies significantly during the daytime, while on weekends the tendency of the level appears to be relatively flat. More interestingly, Figure~\ref{fig:weekdays} shows a complementary relation between the two spaces -- when the mobility level is high in the cyber space, the level is low in the physical space, and vice versa. For example, in the physical space, the mobility level during the work hours (from 9am to 5pm) is relatively low on weekdays. On the contrary, the highest mobility level on weekdays in the cyber space occurs exactly in these hours, indicating that people are more likely to surf the Internet during work hours. The complementary relation can also be observed from the mobility level on weekends -- in the afternoon when the mobility level is high in the physical space, the mobility level in the cyber space decreases correspondingly.  

The temporal properties of human online mobility reveal several interesting relations between the cyber space and the physical space. On one hand, people exhibit the same daily returning pattern, and their mobility levels follow the same day-night cycle in both spaces. This suggests that, although in two different spaces, human circadian rhythm remains unchanged. On the other hand, we discover the complementary relation between the two spaces. This makes sense intuitively -- although sometimes people can access online communities while in transportation, in most cases they cannot ``move'' in both spaces at the same time.


\section{Human Mobility in the Cyber Space}

The similarities between the cyber space and the physical spaces imply the possibility of applying the approaches originally designed for human physical mobility to human online mobility. In this paper, we borrow the idea of entropy-based approaches~\cite{song2010limits}, and study human online mobility from three aspects: 1) the randomness of human online mobility, 2) online mobility patterns, and 3) the preferences to online communities by the people of different mobility patterns. We select entropy-based approaches because such approaches do not require a distance measurement, which is not universally defined in the cyber space.

\subsection{Randomness of Mobility in the Cyber Space}

\begin{figure}
\centering
\includegraphics[width=1\columnwidth]{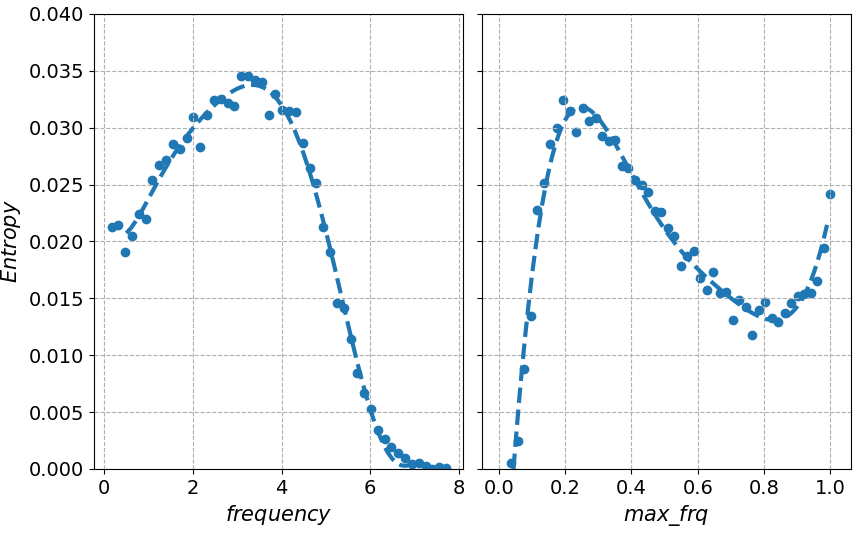}
\caption{Distribution of the entropy of a user's visits to online communities (left). Distribution of the frequency of a user's most visited community $max\_frq$ (right). Please note that the two subplots share the same y-axis.}~\label{fig:entropy}
\end{figure}

Entropy is widely applied to measure the randomness of human mobility in the physical space~\cite{smith2014refined,cranshaw2010bridging}. Inspired by the previous work, we use the entropy of the visit histories of a user to measure the randomness of her mobility across online communities. Formally, the entropy of the visit history of a user $u$, denoted by $En(u)$, is computed as: 
\begin{align}
    En(u) = -\sum_{n} p_{i}log(p_i), 
\end{align}
where $n$ denotes the number of unique communities the user visited, and $p_i$ is the probability of the user visiting the $i$th community. In general, entropy measures how precisely the next community a user will visit can be predicted given the visit history. The lower the entropy is, the lower the mobility randomness is. Another more intuitive measurement that describes the randomness of a user's mobility is the frequency to the most visited community, denoted by $max\_frq(u)$. Formally, $max\_frq(u)$ is computed as
\begin{align}
    max\_frq(u) = max_n(p_{i})
\end{align}
Clearly, a larger value of $max\_frq$ indicates the user is more focused on a specific community, and therefore implies a lower user mobility randomness. To investigate the mobility randomness in the cyber space, we compute both entropy and $max\_frq$ for each user. We conduct the experiments on the same data slice from January to March 2016 in this task, and apply the constraints suggested in~\cite{song2010limits} to remove inactive users (i.e. $n\textgreater2$ and total visits$\textgreater$1000). The distributions of the two measurements are plotted in Figure~\ref{fig:entropy}.

We observe that mobility randomness varies significantly across users. Over 25\% users have an entropy value larger than 4, suggesting that a high randomness of such users' online mobility. Please note that an entropy value of 4 indicates that the next community the user will visit could be found on the average in any of $2^4=16$ communities. Similarly, over 30\% users have a $max\_frq$ value lower than 0.3 indicating that many users do not focus on specific communities. Meanwhile, the mobility randomness of a large portion of users in the cyber space is very low. From the distribution of $max\_frq$, we observe that about 17\% users have a $max\_frq$ value larger than 0.8, and nearly 10\% users with a value larger than 0.9. Please note that a $max\_frq=0.9$ indicates that the user direct 90\% of all the visits to one community. In other words, although these users are active online, they almost always devote their visits to only one community. The distribution of entropy echoes the finding by showing that over 13\% users have an entropy value lower than 1.

\begin{figure*}
\centering
\includegraphics[width=2\columnwidth]{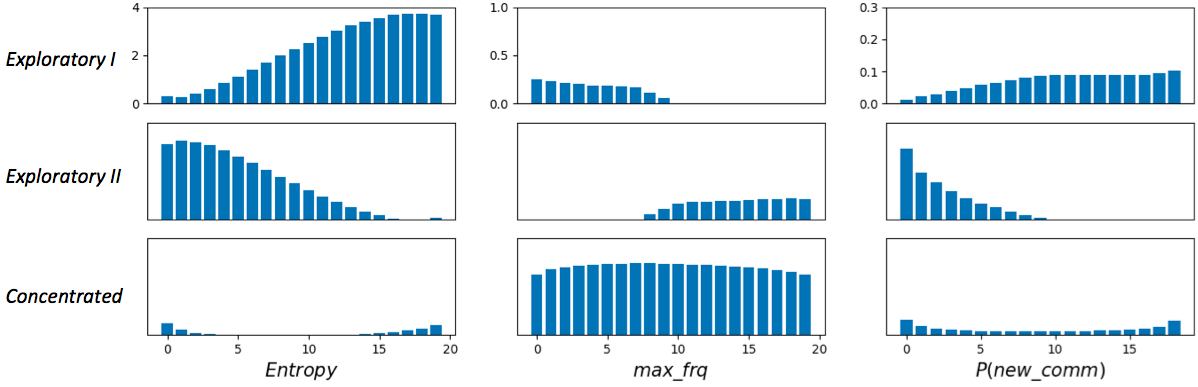}
\caption{Three mobility patterns in the cyber spaces extracted by NMF. For a better illustration, we plot the weights of entropy, $max\_frq$, and $P(new\_comm)$ separately.}~\label{fig:patterns}
\vspace{-1em}
\end{figure*}

\subsection{Mobility Patterns in the Cyber Space}

Given the diverse mobility randomness across people in the cyber space, a nature follow-up question is: does human online mobility change over time? Take the people with very low mobility randomness for an example. We wonder if they always focus on a specific community since they join the platform, or they explore many communities in the beginning and discover their favorites later. To study the problem, we focus on the active users whose whole lifespans on the platform are available~\cite{danescu2013no,tan2015all}. In other words, these users have been active on the platform, but left the platform eventually. Therefore, we select the users who used to be active before January 2016 and never post after the time. We apply the same constraints as in the task for choosing active users (i.e. $n\textgreater2$ and total visits$\textgreater$1000). There are around 69,000 users meeting the constraints. We collect the complete visit histories of these users, and conduct our following experiments on this data slice. 

\subsubsection{Methodology} Since the total number of visits varies significantly across users, we first equally divide the lifespan of a user into 20 stages\footnote{We obtain similar results for other choices of number of stages.} as suggested in~\cite{danescu2013no}. In other words, each stage of a user accounts for 5\% of all the visits through the user's lifespan. By doing this, we are able to align the stages of all the users, and study the evolution of user mobility over different stages. We then quantify the mobility within each stage of a user. Entropy and $max\_frq$ are again used to measure mobility randomness in a stage. However, these two measurements do not quantify to what extent user explore unvisited communities over stages. For example, a user could visit two totally different sets of communities in two stages, but with the same mobility randomness. To measure the user exploration to new community in a stage, we design another measurement $P(new\_comm)$ -- the probability of a user visiting a new community that never has been visited in previous stages. Formally, for the $i$th stage of a user $u$, let $V_{u,i}$ denote the set of the visits of $u$ in this stage. The set of the visits to the communities that are visited in the $i$th stage but never have been visited before is denoted as $V_{u,i}^{new}$. Therefore, $P(new\_comm)$ is computed as
\begin{align}
    P(new\_comm) = \frac{|V_{u,i}^{new}|}{|V_{u,i}|}
\end{align}

For each user, we compute the entropy and $max\_frq$ for all the 20 stages in order. We also compute $P(new\_comm)$ from the second stage to the last stage. We exclude the first stage because all the visits in this stage are directed to new communities, thus the value of $P(new\_comm)$ of the first stage is 1 for all the users. Using the three measurements, we quantify a user's mobility over 20 stages into a 59-dimensional vector. The first 20 dimensions are the entropy values of all the stages, the next 20 dimensions are the $max\_frq$ values, and the remaining 19 dimensions are the $P(new\_comm)$ values of the 19 stages (the first stage is excluded). The vectors of all users are then stacked to create a 59$\times$69,000 matrix. This matrix records the mobility over time for all the users. We investigate the patterns of mobility evolution by decomposing the matrix. Since Non-negative Matrix Factorization (NMF) has been successfully used for mining interpretable temporal-spatial human mobility patterns~\cite{lee1999learning,hu2017tales}, we also apply NMF to complete the decomposition.

\subsubsection{Mobility Patterns} By setting the number of component $k$ to 3, NMF reveals three very interpretable mobility patterns in the cyber space: two exploratory patterns and one concentrated pattern\footnote{With a $k$ value larger than 3, the two exploratory patterns are further decomposed into smaller but less interpretable components, and the concentrated pattern is barely affected. Therefore, we set $k$ to 3 in our experiments}. We plot the three patterns in Figure~\ref{fig:patterns} by showing the tendency of the three measurements over stages. We summarize the patterns as follows:

\begin{itemize}
\item \textbf{\textit{Exploratory Pattern I}}: The people of this pattern concentrate on a few communities in the early stages, but explore more and more new communities with high randomness in the late stages. The entropy value of these users is initially low, and increases over stages. Correspondingly, the $max\_frq$ value starts at a high level, and decreases over stages. The tendencies of both entropy and $max\_frq$ indicate that the mobility randomness is low in the beginning, and goes up as time goes on. In other words, these user only focus on a limited number of communities at first, but gradually lose their concentrations later. Meanwhile, the value of $P(new\_comm)$ is low in the beginning, and increases over stages. This tendency indicates that the users start at a low level of interest in new communities, but explore more and more unvisited communities over stages.

\item \textbf{\textit{Exploratory Pattern II}}: The people of this pattern is the opposite to the first pattern. They explore many new communities in the early stages, but only focus on a few communities later. The mobility randomness is initially high, as indicated by a high entropy value and low $max\_frq$ value. This suggests that these users do not concentrate on any communities in the beginning. As time goes on, the mobility randomness monotonously decreases, as indicated by the decreasing tendency of entropy and the increasing tendency of $max\_frq$ over stages. In other words, the users discover their interested communities, and pay more and more attention to these communities. Meanwhile, the decreasing tendency of $P(new\_comm)$ also suggests that the users pay a large amount of visits to new communities in the early stages, but gradually stop exploring unvisited communities over stages.

\item \textbf{\textit{Concentrated Pattern}}: The people of this pattern are very concentrated -- they direct almost all their visits to a small and unchanged set of communities through the whole lifespan. In this pattern, both entropy value and $P(new\_comm)$ value are low for all the stages. This suggests that these user only visit specific communities, and rarely explore unvisited communities. Correspondingly, the $max\_frq$ is high for all the stages, also indicating the low overall mobility randomness of this pattern.

\end{itemize}


\subsection{Exploratory Patterns vs. Concentrated Pattern}

Given the divergent mobility patterns people exhibit in the cyber space, we further investigate the relation between mobility pattern and online community preference. For example, we wonder if the concentrated type people also like to visit communities that the exploratory people usually visit, and if the people of the two exploratory patterns share similarities in their preferences to communities. We study this problem by converting it to a classification task. In this classification task, we take the three mobility patterns of users as class labels, and attempt to distinguish the three classes only using the communities visited by the users as features. 


\begin{table}
\centering
\begin{tabular}{ | l || c | c | c | }
\hline
&\textbf{\textit{Precision}} & \textbf{\textit{Recall}} & \textbf{\textit{F-1 Score}} \\
\hline
Exploratory I & 0.64 & 0.86 & 0.74 \\
\hline
Exploratory II & 0.32 & 0.10 & 0.15\\
\hline
Concentrated & \textbf{0.77} & \textbf{0.71} & \textbf{0.74}\\
\hline
avg / total & 0.63 & 0.67 & 0.63\\
\hline
\end{tabular}
\caption{Classification results among the three patterns.}~\label{tab:classification}
\vspace{-1em}
\end{table}

\subsubsection{Classification among Three Mobility Patterns} 

From the results of NMF, we find the mobility pattern to which a user is assigned the highest weight among the three patterns, and take the mobility pattern as the user's class label. By doing so, we obtain around 31,000 and 13,000 users of Exploratory Pattern I and Pattern II, respectively. The around 25,000 remaining users are labelled as the Concentrated Pattern. The amounts of visits to different communities of a user are used as the classification features. To be specific, we first remove the unpopular communities that have less than 50 unique users, and obtain 7,526 unique communities. We collect the numbers of visits to the selected communities for each user. TF-IDF is then applied to weight the numbers of visits across all the users, and the weighted results are used as the classification features. We use 80\% of the data for training, and the remaining 20\% for testing. A logistic regression model is employed for the task. The classification results are reported in Table~\ref{tab:classification}.

The results show that people of the Concentrated Pattern can be distinguished from the people of Exploratory Pattern I and Pattern II, with a high precision of 0.77 and recall of 0.71. This indicates that concentrated people have different preferences to communities from exploratory people. However, the classifier cannot distinguish people of Exploratory Pattern I and II. Because of the much larger user size of Exploratory Pattern I, a large amount of users of Exploratory Pattern II are classified as Exploratory Pattern I. This leads to the low precision (0.32) and recall (0.1) for Exploratory Pattern II. Due to the same reason, Exploratory Pattern I receives a high recall (0.86) but a low precision (0.64). The results imply that users of the two exploratory patterns share similarities in their community preferences, and therefore cannot be simply distinguished by only using the community features.

\begin{table}
\centering
\begin{tabular}{  | l | c | c| }
\hline
\textbf{\textit{Subreddit}} & \textbf{\textit{User Size}} & \textbf{\textit{Coefficient}}\\
\hline
\multicolumn{3}{|c|}{\textit{top five positive features (communities)}}\\
\hline
/r/stopdrinking & 94,187 & 0.013 \\
\hline
/r/thinkpad & 24,545 & 0.012\\
\hline
/r/MinecraftCirclejerk & 2,508 & 0.011\\
\hline
/r/incremental\_games & 39,250 & 0.011\\
\hline
/r/autism & 21,461 & 0.011\\
\hline
\multicolumn{3}{|c|}{\textit{top five negative features (communities)}}\\
\hline
/r/bestof & 4,841,958 & -0.083\\
\hline
/r/reactiongifs & 1,260,843 & -0.064\\
\hline
/r/woahdude & 1,571,722 & -0.051\\
\hline
/r/comics & 836,383 & -0.050\\
\hline
/r/technology & 5,866,352 & -0.041\\
\hline
\end{tabular}
\caption{Five top positive and negative features (communities) for classifying the people of the Concentrated Pattern. We list the communities along with their user sizes and the values of the coefficients.}~\label{tab:features}
\vspace{-1em}
\end{table}

\subsubsection{Community Preference of Different Patterns} 

From the trained logistic model, we can tell the features (communities) assigned with the highest positive and negative coefficients for distinguishing the users of the Concentrated Pattern. These communities are reported in Table~\ref{tab:features}. Clearly, the communities with the positive coefficients are preferred by the users of the Concentrated Pattern. In contrast, the communities with the negative coefficients are preferred by the users of either Exploratory Pattern I or Pattern II. We also report the values of the coefficients and the user sizes of the communities in the table. Two interesting differences can be observed from these two groups of communities. First, the communities preferred by the concentrated type people are much smaller than the communities preferred by the exploratory type people. None of the top five positive communities has a user size larger than 100,000, and the smallest only has a user size of 2,500. In contrast, four of the top five negative communities have a user size larger than 1 million. Furthermore, the communities preferred by the concentrated type users are usually on specific topics. For example, in the top five positive communities, two communities are on specific games (/r/MinecraftCirclejerk and /r/incremental\_games), two communities are on personal issues (/r/stopdrinking and /r/autism), and one community is on a specific product (/r/thinkpad). In contrast, the top five negative communities are all on general topics such as /r/bestof and /r/reactiongifs.

\section{Conclusion \& Future Work}

In this paper, we present a novel analogy between the cyber space and the physical space. We quantitatively validate the analogy from three representative and progressive aspects: visit distributions, individual visit behavior, and temporal properties. Our experiments on the three aspects all reveal striking the similarities between human mobility in the two spaces. Next, we study human online activities by treating the communities as locations in the cyber space, and activities across communities as movements across locations. By applying the framework originally designed for studying human physical mobility, we investigate the mobility patterns in the cyber space. It is observed that the randomness of online mobility varies significantly across users. Furthermore, we study the evolution of human mobility in the cyber space, and discover three interesting exploration patterns of by users of online communities. Moreover, our experiments suggest divergent preferences to different online communities across people of different patterns. Our work provides valuable insights into the human activities under the multi-community setting. More importantly, we uncover the interesting similarity between the two spaces, and suggest a promising research direction.

In the future, we plan to build upon our work mainly from two aspects. First, we would like to quantify the ``cost'' of users moving among communities, i.e. the ``distance'' in the cyber space. Such a distance measurement would allow us to apply more well-established frameworks for modelling the physical movements to the online scenarios. Second, we would like to study online communities by borrowing the approaches developed for studying physical locations. Much previous work discusses the characteristics of physical locations. It would be interesting to see if the locations in the cyber space exhibit similar characteristics.

\bibliographystyle{aaai}

\bibliography{reference}

\end{document}